# An improved design of spark-protected microstrip gas counters (R-MSGC

R. Oliveira,[1] V. Peskov,[1,2] F. Pietropaolo,[3] P.Picchi[4]
[1]CERN, Geneva, Switzerland
[2]UNAM, Mexico
[3]INFN Padova, Padova, Italy
[4]INFN Frascati, Frascati, Italy

**Abstract**
We have developed microstrip gas counters manufactured on standard printed circuit board and having the following features: resistive cathode strips, thin (10μm) metallic anode strips and electrodes protected against surface discharges by a Coverlay layer at their edges. These features allow the detector to operate at gas gains as high as can be achieve with the best microstrip gas counters manufactured on glass substrates. We believe that after further developments this type of detectors can compete in some applications with other micropattern detectors, for example MICROMEGAS.

## I. Introduction

Nowadays, micropattern gaseous detectors (MPDGs) challenge traditional gaseous detectors (such as MWPCs, RPCs and parallel-plate avalanche chambers) in practically all applications. The main advantage of the MPGDs is that they are manufactured by means of microelectronics technology, which offers high granularity and consequently an excellent position resolution. However, the fine structure of their electrodes and the small gap between them make MPGDs electrically "weak." In fact, their maximum achievable gains are usually not very high and they can be easily destroyed by sparks, which may occur during their operation.

A few years ago, we developed the first GEM-type micropattern detectors featuring resistive electrodes instead of metallic ones [1]. The resistive electrodes limit the current during the sparks and make the detector, as well as the front-end electronics, spark-protected. This work triggered a sequence of similar developments, which are nowadays performed not only by our group, but by several other groups in the frame work of CERN RD51 collaboration [2]. Examples could be: resistive strip RETGEM [3], resistive CAT/WELL detector [4-6]] and resistive mesh detectors [7]. The most significant among these works was the successful development and tests of large-area MICROMEGAS with resistive electrodes [8].

In the light of these developments we recently started investigating whether microstrip gas counters (MSGCs) can also be made spark-protected. As discussed in the conclusion section of this paper this type of detector can be attractive for some applications.
We have already reported the results of tests of two prototypes of MSGCs with resistive electrodes (we named the R-MSGCs) [2, 9, 10].
The first one had both the anode and the cathode strips manufactured on a printed board circuit. The maximum achievable gain of this R-MSGC was not high ($A_m \sim 10$) [2], however it was demonstrated that such a detector is fully spark-protected. The second prototype had resistive cathodes and metallic anodes and allows to reach much higher gains- close to 500 [9].

In this work for the first time we successfully applied a modified technology of the R-MSGC manufacturing. This allowed high quality detectors to be produced which were capable of operating at gas gains as high as those achieved by the best "classical" MSGCs with metallic electrodes on glass substrate.

## II. Detector manufacturing

The new detector was manufactured from a PCB 2.2 mm thick the top surface of which was coated with 5 μm thick Cu layer. Then on the top surface of the PCB parallel grooves were milled 100 μm deep, 0.6 mm wide, with pitch 1mm (see Fig.1). The grooves were then filled with resistive paste (ELECTRA Polymers) and the R-MSGC surface was chemically cleaned. Finally by using a photolithographic technology Cu 20 μm wide strips were created between the grooves. This technology is simpler than we used before for manufacturing prototypes #1 and 2 [2, 9] and consequently the cost of the detector is lower.

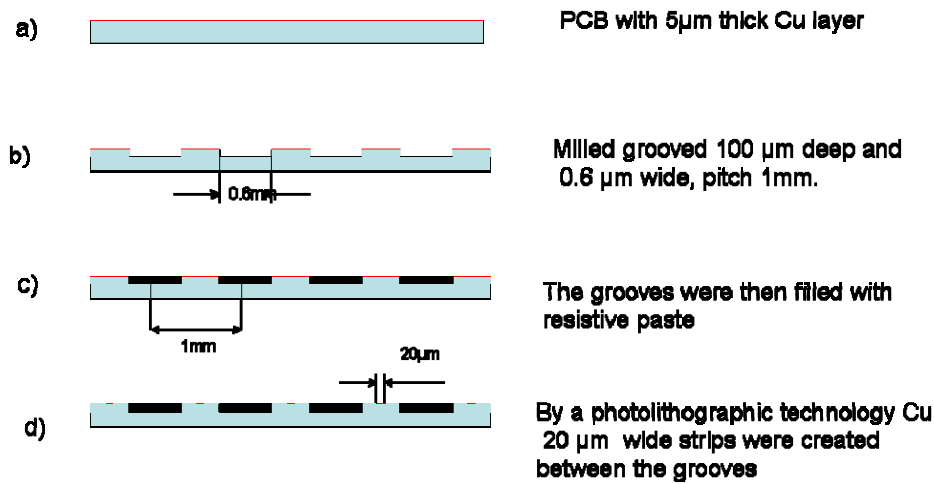

Fig.1. Schematic illustration of a R-MSGC manufacturing process

Both the anode and the cathode strip were covered near their edges by a Coverlay layer to avoid surface discharges (see Fig.2). The total resistivity of each cathode strips was 300 MΩ and their resistivity in the region C (see Fig. 2) was 100MΩ.

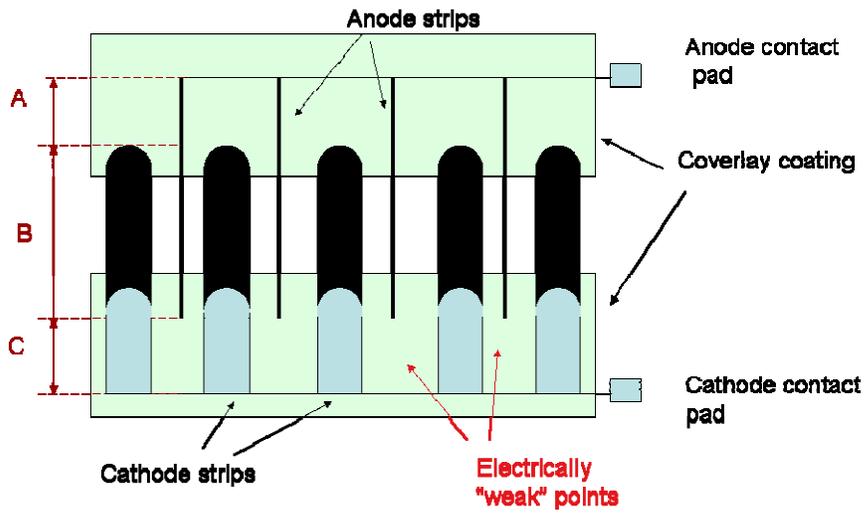

Fig. 2. A schematic drawing of the R-MSGC

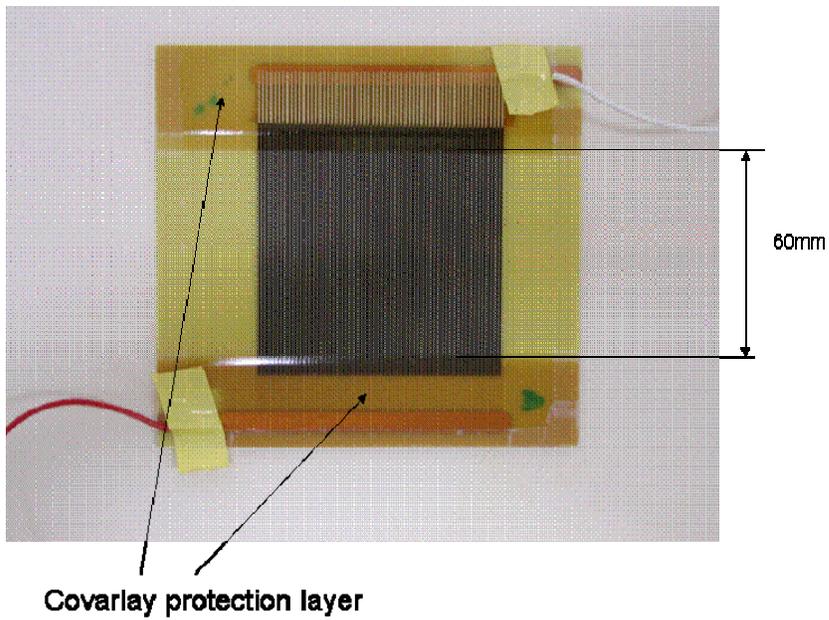

Fig. 3. A photograph of the R-MSGC

The photograph of the R-MSGC is presented in Fig. 3 whereas a magnified photo of the anode and the cathode strips is seen in Fig. 4.

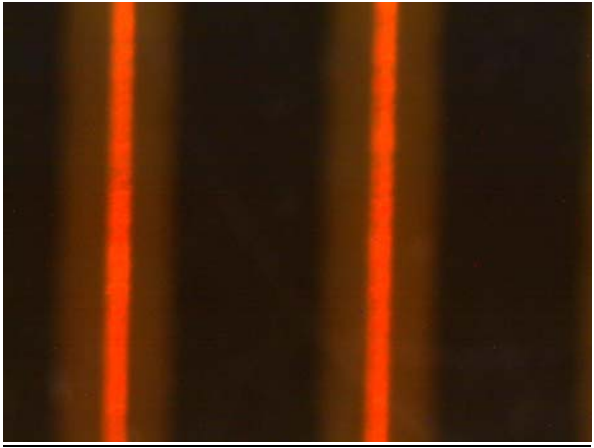

Fig. 4. A magnified photo of the central region of R-MSGC showing metallic anode strip (red) and resistive cathode strips (black)

The new R-MSGC thus has the following three features:
1) very this metallic anode strips, 2) resistive cathode strips manufactured by filling grooves with a resistive paste, 3) a Coverlay layer to protect the edges against surface discharges. These three features allow the detector to operate at gas gains much higher than could be achieved with previous R-MSGCs prototypes [2, 9].

### III. Results

The detector characteristics were measured using the setup described in [9]. Tests were performed in Ne and Ar and their mixtures with $CH_4$ and $CO_2$ in the range of a quencher concentration between 10 and 20%.
Figures 5-7 show curves of R-MSGC gain vs. voltage measured in Ne and Ar and in some of their mixtures with $CH_4$ and $CO_2$. As can be seen in all these gases and mixtures the maximum achievable gain was $10^4$. This gas gain is as high as typically can be achieved with the best MSGC manufactured on glass substrate.
The energy resolution measured in Ar+$CH_4$ and Ar+$CO_2$ gas mixtures was about 25% FWHM for 6keV photons; in mixtures with Ne it was about 28%
The gas gain variations measured as a function of the counting rate at a gas gain of $510^3$ are shown in Fig 8. As can be seen the signal amplitudes started dropping at counting rates of more than 100Hz/mm$^2$ and this was mainly due to the substrate surface charging up since the resistivity of the strips did not create a noticeable voltage drop at this avalanche current.

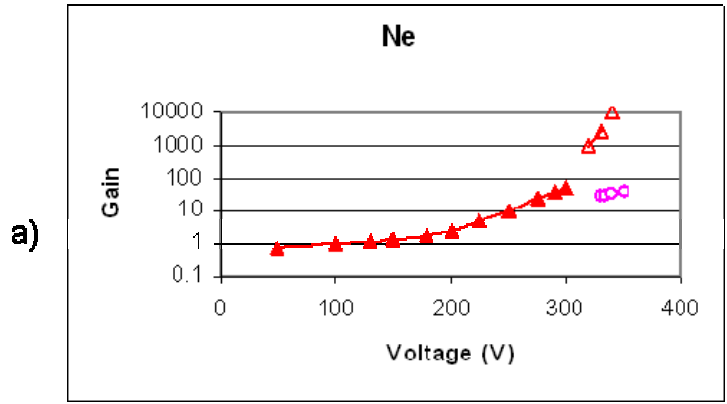

a)

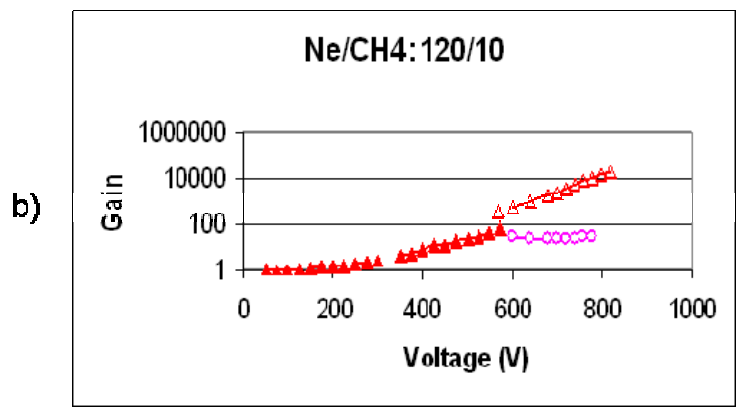

b)

Fig.5. Gas gain vs. the voltage of R-MSGC measured in Ne (a) and Ne+7%CH$_4$ (b) with alpha particles (filled triangles) and with $^{55}$Fe (empty triangles). The curves with circles represent the energy resolution (FWHM) at 6keV

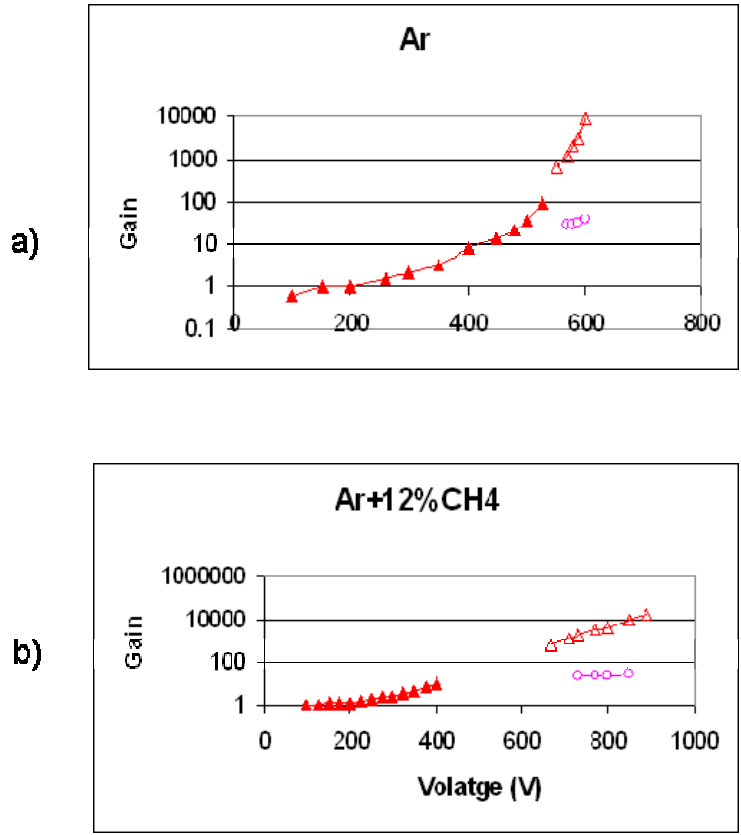

Fig. 6. Gain (triangles) and energy resolution (circles) dependence on voltage applied to R-MSGC measured in Ar (a) and Ar+12%CH$_4$ (b). Filled triangles-measurements performed with alpha particles, open triangles - $^{55}$Fe.

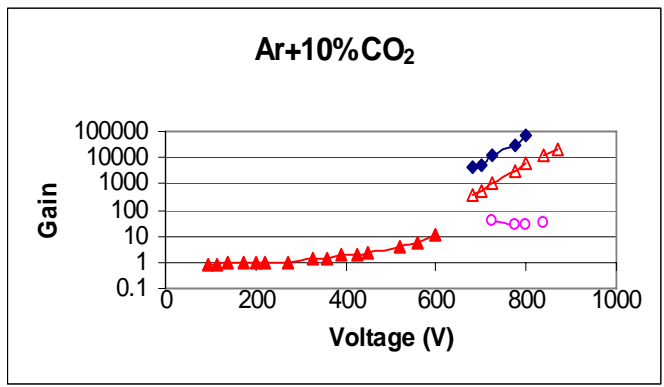

Fig.7. Gas gain curves measured in Ne+10%CO$_2$: filled triangles –measurements performed with alpha particles, open symbols- measurements with $^{55}$Fe. Rhombuses represent gas gain measures with a preamplification structure –resistive meshes [7]. Open circles-energy resolution

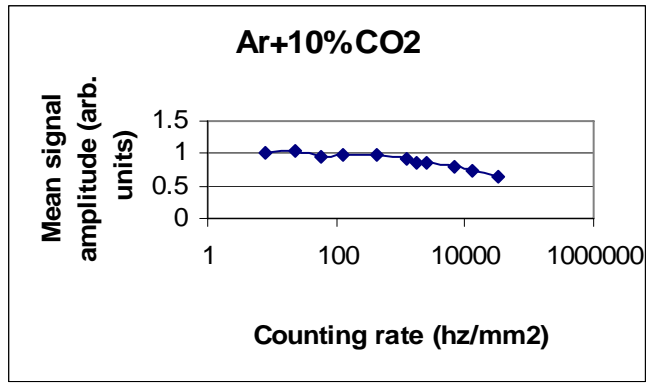

Fig. 8. The gas gain variations with counting rate. Measurements were performed in Ne+10%CO$_2$ at gas gain of 510$^3$.

## IV. Discussion

As was shown in our earlier works [11-13], the maximum achievable gain in MSGC is usually limited by the surface streamers. Surface streamers may prevent the detector from reaching the Raether limit [14] so the gas gain achieved with 6keV photons of the best MSGC is typically around 10$^4$ or lower. The most efficient way of streamer suppression is to use very thin anode strips and of course the detector substrate should be very clean and smooth (this is why glass substrate is usually the best). However, the earlier attempts of different authors to implement narrow strips were not very successful; they usually were easily burned by discharges appearing at high gains
In this work we have succeeded in implementing a thin anode strip approach: sparks appearing at gains >10$^4$ were not able to destroy these thin anode strips due to the resistivity of the cathode. Another important feature of our R-MSGC was the Coverlay coating of the electrodes near their edges which prevented surface streamers. By incorporating these two features (resistive coating and Coverlay coating) the detector could operate at gas gains as high as one can achieve with MSGC manufacture on a glass sub stare. The surprise was that cleaned printed circuit surfaces can provide the same discharge strength (a threshold for the appearance of surface streamers) as well -cleaned glass surfaces.
Note also that gains achieved with R-MSGCs (~10$^4$) are comparable to those obtained with spark-protected MICROMEGAS having resistive anode strips (R-MICROMEGAS)- see Fig.9 reproduced from [8]) .The rate characteristics of R-MSGCs are as good as those of R- MICROMEGAS –see Fig.10 [8].

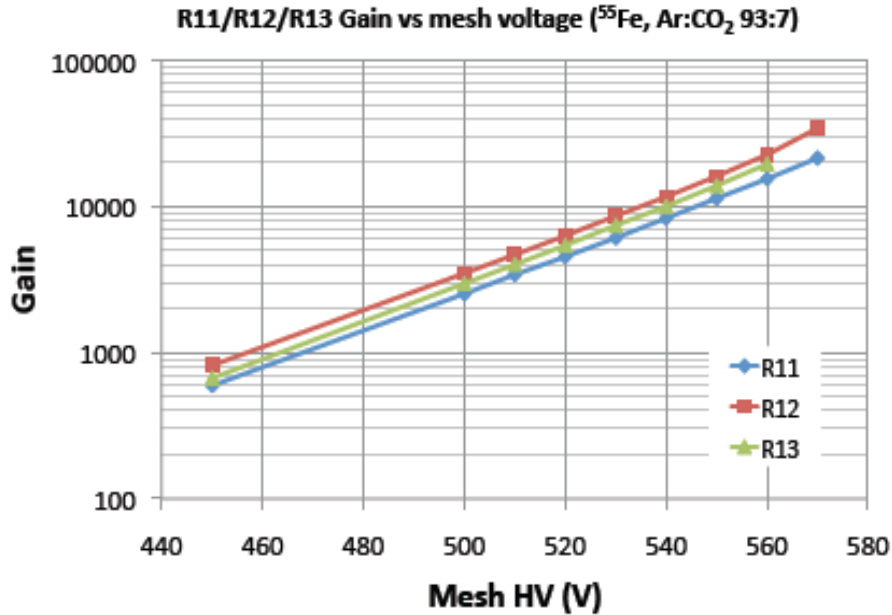

Fig.9. Gas gain curves measures with R-MICROMEGAS [8].

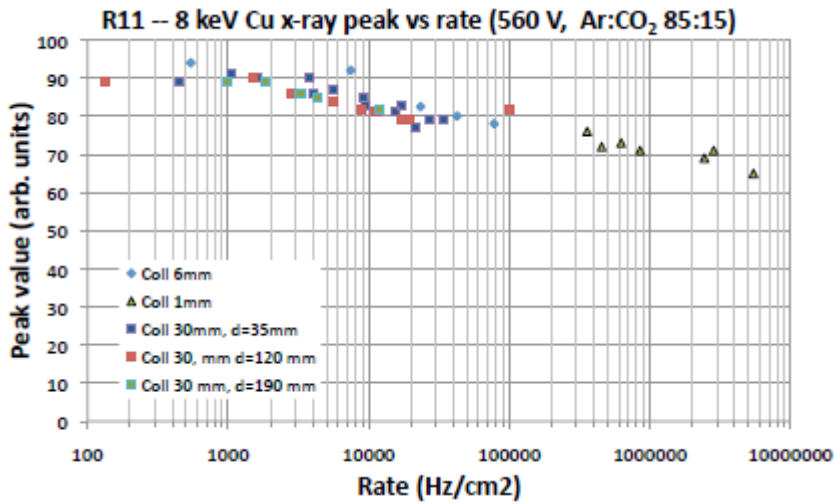

Fig.10. Rate characteristics of R-MICROMEGAS [8]

However, R-Mugs have several advantages over R-MICROMEGAS:
1) R-MSGCs are easier to manufacture than MICROMEGAS,
2) It is easier to clean the form dust particles (since there is no cathode mesh which blocks these microparticles)
3) There are fewer parts in R-MSGCs than in R-MICROMEGAS
4) Large area R-MSGCs can be assembled from patches with practically no dead spaces (in contrast to R-MICROMEGAS)
Thus R-MSGCs appear to be a very attractive and competitive detector of photons an

## V. Work in progress/ future plans

As was shown in earlier works (see for example [15-17]) MSGC have excellent 2-D position resolution.
We have already built a prototype of 2-D sensitive R-MSGC a photo of which is shown in Fig. 11. In the centre of this photo one can see the active area of the R-MSGC: its anode and cathode strips. On the right side of the photo and on its bottom are shown the rows of readout strips located under the R-MSGC (similar to the R-MICROMEGAS design); one row placed parallel and the other one perpendicular to the strips of R-MSGC. The preliminary results indicate that the position resolution ~ 100 µm can be achieved by applying a center of gravity method to the measured indices signals from these strips.

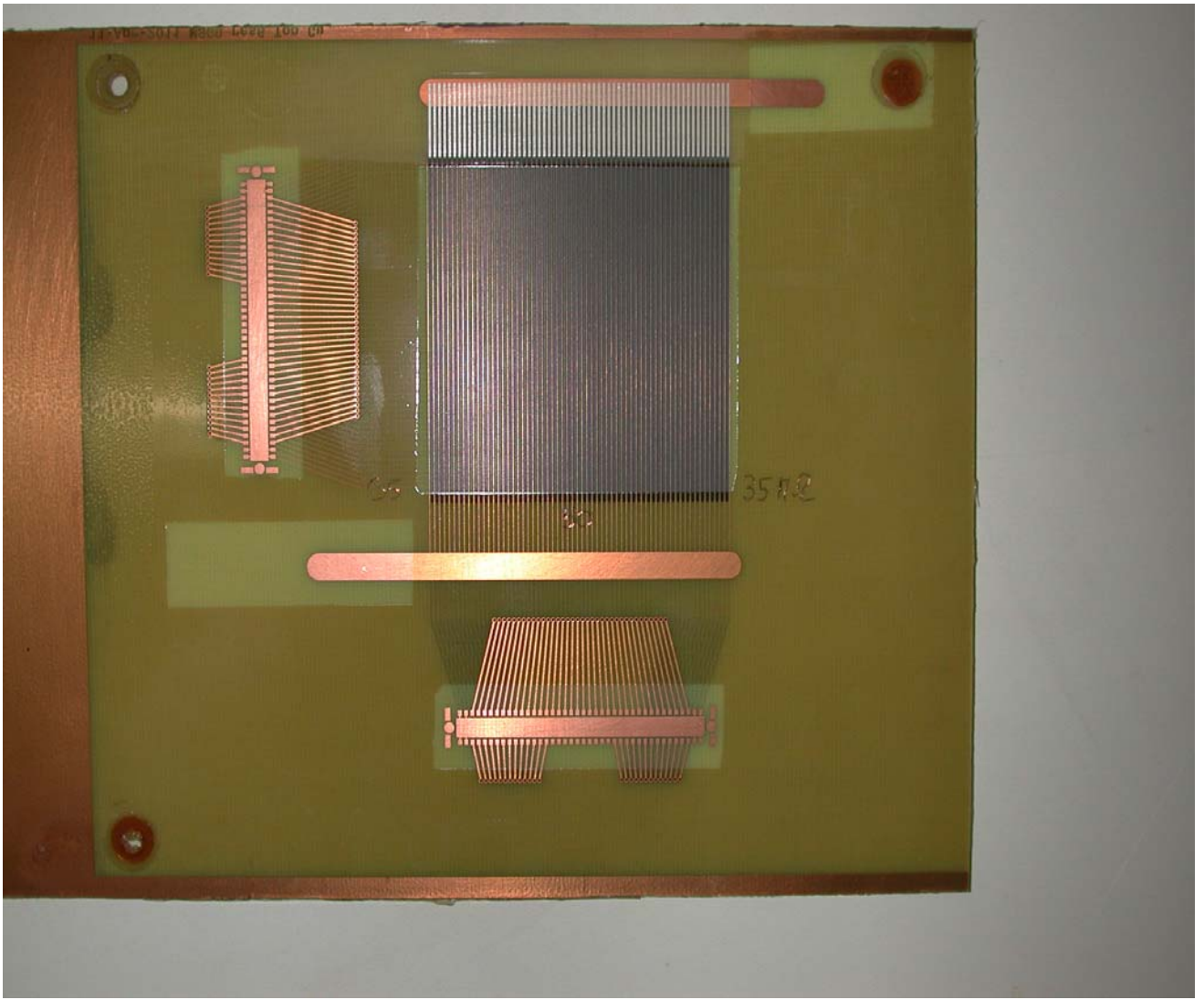

Fig.11. A photo of a 2-D sensitive R-MSGC having capacitor coupling readout strips

## VI. Conclusions.

Our progress in developing R-MSGCs is reflected in Table 1 (see earlier papers [1, 2] for description of prototypes #1 and #2). In this sequence of prototypes each has a narrower anode width than the previous one. The breakthrough in gas gain achieved with the present prototype #3 was due to the thin anode strips.

| R-MSGC -type | Max. achievable gain with 6keV photons |
|---|---|
| 1 | ~10 |
| 2 | ~500 |
| 3 | 10000 |

Table.1 Maximum achievable gains of various R-MSGC prototypes

In conclusion the maximum achievable gain of an R-MSGC is equal to that achieved with the best glass MSGC and approaching the maximum achievable gain of R-MICROMEGAS. However, a R-MSGC is easier to manufacture than MICROMEGAS and easier to clean from dust particles. R-MSGCs have fewer parts in the detector assembly (no cathode mesh) and can be assembled from patches with practically no dead spaces/zones
We believe hat after further developments R-MSGCs can compete in some applications with other micorpattern gaseous detectors, for example MICROMEGAS

## VII . References: